\documentclass[aps,prl,twocolumn,preprintnumbers,groupedaddress,superscriptaddress,floatfix,tightenlines,reprint]{revtex4-1}
\usepackage{mathrsfs,natbib}
\usepackage{graphics,epsfig,color, graphicx}
\usepackage{verbatim}

\usepackage{graphicx,epsf,amssymb,bbm,amsbsy,amsfonts,amssymb,amsmath}
\usepackage{hyperref}

\def\Th#1#2{\vartheta\left[\,{}^{#1}_{#2}\,\right]}

\newcommand{\eq}{\begin{equation}} 
\newcommand{\eqe}{\end{equation}}

\newcommand{\eqa}{\begin{eqnarray}}
\newcommand{\eqae}{\end{eqnarray}}

\newcommand{\im}{\mathrm{Im}\,}
\newcommand{\re}{\mathrm{Re}\,}

\newcommand{\ii}{\mathit{i}}

\def\Th#1#2{\vartheta{\tiny\begin{bmatrix}
{#1}\\
{#2}
\end{bmatrix}}}

\def\half{{\textstyle{1\over 2}}}

\hypersetup{
    colorlinks=true,       % false: boxed links; true: colored links
    linkcolor=red,          % color of internal links (change box color with linkbordercolor)
    citecolor=blue,        % color of links to bibliography
    filecolor=magenta,      % color of file links
    urlcolor=blue           % color of external links
}

\begin{document}
%%%%%%%%%%%%%%%%%%%%%%%%%%%%%%%%%%%%%%%%%%%%%%%%%%%%%%%%%%%%%%%%%%%%%%%%%%%%%%%%%%%%%%%%%%%%%%%%%%%%%%%%%%%%%%%%%%%%%%%%%%%%%%%%%%%%%%%%%%%%%%%%%%%%%%%%%%%%%%%%%%%%%%%%%%%%%%%%%%%%%%%%%%%%%%%%%%%%%%%%%%%%%%%%%%%%%%%%%%%%%%%%%%%%%%%%
%%%%%%%%%%%%%%%%

\title{A 4D-2D equivalence for large-$N$  Yang-Mills theory}

%\preprint{XZZZZZ}
%  \preprint{FTPI-MINN-15/35, UMN-TH-3444/15, IPMU-{\bf XXXX}}

\author{G\"ok\c ce Ba\c sar}  
\email{gbasar@umd.edu}
\affiliation{%Maryland Center for Fundamental Physics, 
Department of Physics,
University of Maryland, College Park, MD 20742  USA}

\author{Aleksey Cherman}
\email{acherman@umn.edu}
\affiliation{%Fine Theoretical Physics Institute, 
Department of Physics, University of Minnesota, Minneapolis, MN 55455 USA}

\author{Keith R.\ Dienes}
\email{dienes@email.arizona.edu}
\affiliation{Department of Physics, University of Arizona, Tucson, AZ 85721 USA} 
\affiliation{%  Maryland Center for Fundamental Physics, 
Department of Physics,
University of Maryland, College Park, MD 20742  USA}

\author{David A.\ McGady}
\email{david.mcgady@ipmu.jp}
\affiliation{Kavli IPMU (WPI), University of Tokyo, Kashiwa, Chiba 277-8586, Japan}
\affiliation{Department of Physics, Princeton University, Princeton NJ 08544 USA}

\begin{abstract}
General string-theoretic considerations suggest that four-dimensional large-$N$ gauge theories should have dual descriptions in terms of two-dimensional conformal field theories. However, for non-supersymmetric confining theories such as pure Yang-Mills theory, a long-standing challenge has been to explicitly show that such dual descriptions actually exist.  In this paper, we consider the large-$N$ limit of four-dimensional pure Yang-Mills theory compactified on a three-sphere in the solvable limit where the sphere radius is small compared to the strong length scale, and demonstrate that the confined-phase spectrum of this gauge theory coincides with the spectrum of an irrational two-dimensional conformal field theory.
\end{abstract}


\maketitle

%%%%%%%%%%%%%%%
 {\bf Introduction.}
Confining gauge theories in the large-$N$ limit are believed to have dual descriptions as weakly-coupled string 
theories~\cite{tHooft:1973jz,*Witten:1979kh}.  Since string theories have 2D worldsheet 
conformal field theory (CFT) descriptions, it is expected that confining 4D gauge theories may have alternative descriptions based on 2D CFTs.  However, for non-supersymmetric quantum field theories (QFTs) such as Yang-Mills (YM) theory, no concrete 
relation between large-$N$ confining theories and 2D CFTs has ever been found.

\begin{figure}[t]
  \centering
\includegraphics[width=0.47\textwidth]{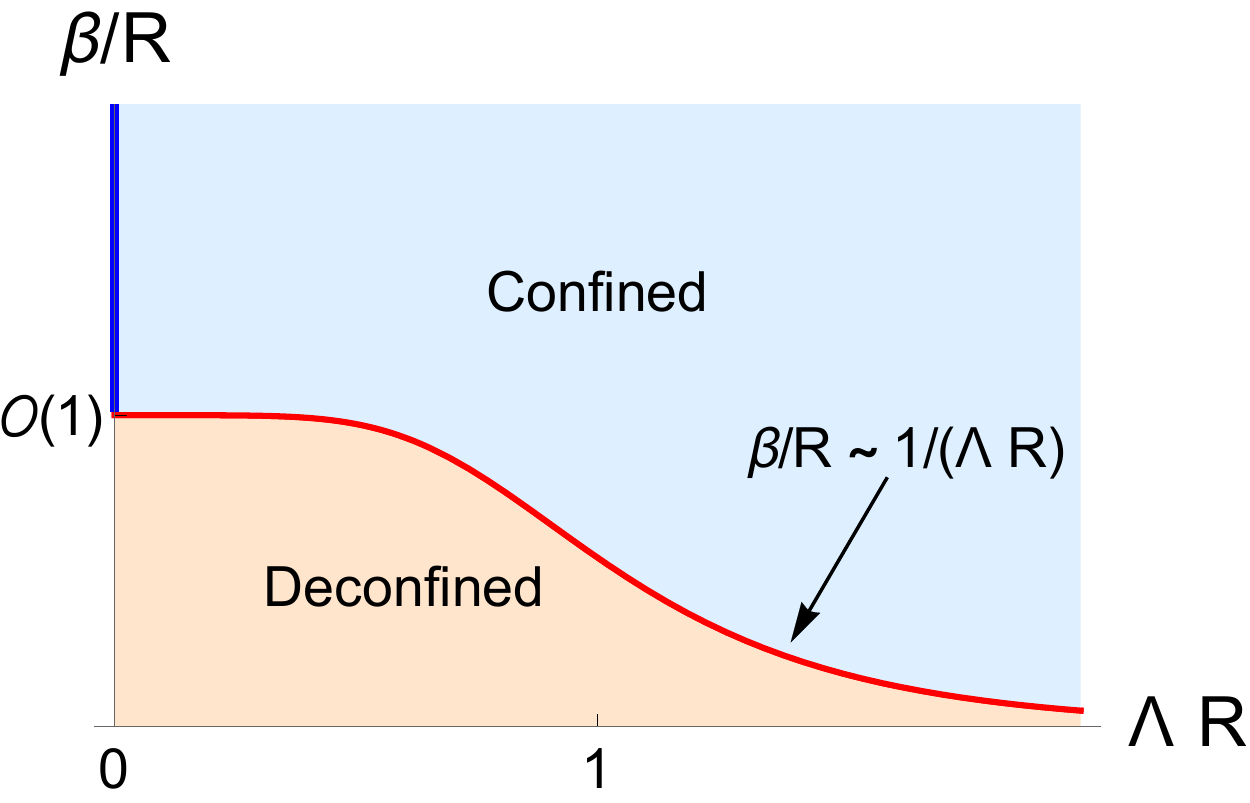}
  \caption{A conjectured phase diagram for large-$N$ YM theory on $S^3 \times S^1$.
In the analytically tractable regime $\Lambda R \ll 1$, the deconfinement transition occurs
at $\beta \sim R$, while for $\Lambda R \gg 1$, lattice studies have shown
that it occurs at $\beta \sim 1/\Lambda$.
This sketch illustrates the natural conjecture that these two limiting cases are smoothly connected.
The results of this paper apply in the $\Lambda R \to 0$ region indicated by the blue line.
} 
\label{fig:YMPhaseDiagram}
\end{figure}

In this paper we tackle this problem by studying the large-$N$ 
limit of 4D pure $SU(N)$ YM theory, formulated at 
temperature $T  = \beta^{-1}$ 
and compactified on a three-sphere $S^3$ of radius $R$.
One can thus view the theory as living on $S^3_R \times S^1_{\beta}$ with Euclidean metric signature.     
The virtues of this setting are two-fold.
First, thanks to asymptotic freedom, if we take $\Lambda R \ll 1$ where $\Lambda$ is the YM strong 
scale,
then the 't Hooft coupling $\lambda \equiv g^2  N$ becomes small --- {\it i.e.}\,  $\lambda(1/R)\to 0$.
As a result, the theory becomes solvable for any temperature $\beta \sim N^0$.  
Second, it is known~\cite{Aharony:2003sx} that large-$N$ YM theory stays in the confined phase 
when $\beta/R \gtrsim 1$, even when $\lambda \to 0$. In this context 
``confinement'' means that the system has an unbroken center 
symmetry and that its free energy scales as $N^0$.  
As sketched in Fig.~\ref{fig:YMPhaseDiagram},
it is plausible that the physics of YM theory is smooth as a function
 of $\Lambda R$.  Thus, the $\Lambda R \ll 1$ regime of the large-$N$ confined phase represents a particularly tractable 4D starting point in our search for a dual 2D description. 

%%%%%%%%%%%%%%
%%%%%%%%%%%%%%

Rather than attempt a string-theory construction of a 2D dual %description 
for large-$N$ YM theory, we 
shall instead analyze the confined-phase spectrum of YM theory in the solvable $\Lambda R \ll 1$ limit.  Remarkably, we find that
a simple 2D CFT description emerges.   
Thus, in this limit, we conclude that the large-$N$ confined-phase spectrum of 4D YM theory coincides with the spectrum of a 2D CFT.  In the conclusions we briefly comment on possible relations between our result and string-theoretic expectations.

Specifically, recall that 
the complete spectrum of a QFT is encoded in its grand-canonical thermal partition function.
We take 4D YM theory to be minimally coupled to the $S^3$ metric, so that the Kaluza-Klein energies on the three-sphere are given by $E_n=n/R$ in the $\lambda \to 0$ limit~\cite{Aharony:2003sx}.  The partition function then takes the form
\begin{equation}
  Z_{\rm YM}(\beta/R) ~=~ \sum_{n=0}^{\infty} d_n  e^{-\beta E_n} ~=~ \sum_{n=0}^\infty d_n q^n~ 
\label{eq:PartitionSumForm}
\end{equation}
where $q = e^{-\beta/R}$ and $d_n$ counts the number of states with energy $E_n$. Our main result is the demonstration that the grand-canonical partition function $Z_{\rm YM}$ of Yang-Mills theory 
coincides with a chiral partition function of a 2D CFT:
\begin{equation}
    Z_{\rm YM}(\tau) ~=~ 
            Z_{\rm 2D}(\tau) ~.
 \label{eq:TheClaimYM}
\end{equation}
In writing Eq.~\eqref{eq:TheClaimYM}, we have
analytically continued $q$ to $e^{2 \pi i \tau}$ with $\tau \in \mathbb{H}$, the complex 
upper half-plane.  Thus $\im \tau = \beta/(2\pi R)$.  
Determining the physical interpretation of $\re \tau$ on the 4D gauge-theory 
side of Eq.~\eqref{eq:TheClaimYM} is an important problem for future work.

%%%%%%%%%%%%%%%%%%%%%%%%%%%
{\bf The 4D partition function.}  We begin by briefly explaining the computation of $Z_{\rm YM}$, leaving a more leisurely exposition to Ref.~\cite{GCDM}.  To calculate the 4D partition function $Z_{\rm YM}(\tau)$,
we take the large-$N$ limit with $\Lambda$ held fixed, which means taking the continuum limit after the large-$N$ limit.    We work on $S^3 \times S^1$ and assume that $\beta$ and $R$ are independent of $N$.   Likewise, we do not consider states with energies $\gtrsim N$ because they lie beyond our UV cutoff.  As is typical in studies of large-$N$ theories, we work with the $U(N)$ version of YM theory rather than the $SU(N)$ version~\footnote{In YM theory the overall $U(1)$ in $U(N)$ decouples, so its contribution to the partition function factorizes.}.  When $\Lambda R \to 0$, the microscopic degrees of freedom of YM theory reduce to an infinite collection of color-adjoint-valued harmonic oscillators. These oscillators are counted by the massless-vector partition function, which can be written as $z_v(\tau) = (6q^2 - 2q^3)/(1-q)^3$. The physical states are then determined by imposing the color Gauss law.  In the $\lambda = 0$ confined phase, the physical single-particle states can be identified with single-trace operators, and their energies are proportional to their scaling dimensions.  The counting problem for these states, and also for the multi-particle states, has been solved~\cite{Aharony:2003sx,Sundborg:1999ue,*Polyakov:2001af}, and the resulting grand-canonical confined-phase partition function is given by
\begin{align}
   Z_{\rm YM}(\tau) &= \prod_{n=1}^{\infty} \frac{1}{1-z_v(q^n)} = \prod_{n=1}^{\infty} \frac{(1-q^n)^3}{1-3q^n -3 q^{2n} +q^{3n}} \nonumber\\
    &= 1 +  6 q^2 + 16 q^3 + 72 q^4 + ...
\label{eq:YM1}
\end{align}

As expected in any confining large-$N$ theory, 
we find that the $d_n$ grow exponentially for large $n$. Thus, there are Hagedorn singularities in $Z_{\rm YM}(\tau)$.
In Eq.~(\ref{eq:YM1}), we find $d_n\sim e^{C n}$ and $ E_n \sim n$ for large $n$, with  
$C\equiv \log(2+\sqrt{3})\approx 1.317$. This contrasts with the behaviors
$d_n \sim e^{\sqrt{n}}$ and $E_n \sim \sqrt{n}$ that would arise 
for a string theory with a flat target space. 
Of course, we are not in flat space:  the spacetime curvature is $\sim 1/R$, which is of the same scale as the effective string tension $\alpha' \sim 1/R^2$ that follows from our spectrum.
The scaling properties of $d_n$ in Eq.~\eqref{eq:YM1} imply that the leading Hagedorn singularity of $Z_{\rm YM}(\beta, \mu_I)$ is at $\beta_H/ R = C, \mu_I = 0$.
Consequently, at $\mu_I=0$, there must be a phase transition to a deconfined phase 
at $\beta_H$. This is discussed in detail in Refs.~\cite{Aharony:2003sx,Aharony:2005bq}.
% for extensive discussions. 

%%%%%%%%%%%%%%%%%%%%%%%%%%%
{\bf Modular symmetries.}
We now observe that the denominator in Eq.~\eqref{eq:YM1}
can be factorized with roots that are inverses of each other:
\begin{equation}
   1-3q^n -3 q^{2n} +q^{3n} ~=~ (1+q^n) (1-q^n z) (1-q^n/z)
\end{equation}
where $z=2+ \sqrt{3}$. %and $w = 1/z =  2- \sqrt{3}$.  
This pivotal algebraic observation was first made in Ref.~\cite{Basar:2014mha}  in the context of uncovering a subtle ``temperature-reflection'' symmetry for $Z_{\rm YM}$.
For our purposes, however,  the key point is that this allows
%$w=1/z$ implies that
$Z_{\rm YM}$ to be written as %in the form
%For our purposes, however, the key point is that the fact that $w=1/z$ implies that $Z_{\rm YM}$ can be written in the form
\begin{align}
    Z_{\rm YM} &= \prod_{n=1}^{\infty}\frac{(1-q^{n})^3}{(1+q^n)(1-q^n z)(1-q^n z^{-1})}~.
\label{eq:YM3}
\end{align}

This observation is very important because the structure of Eq.~\eqref{eq:YM3} 
matches the structure of the product 
representations of the Dedekind $\eta$-function  and
generalized Jacobi $\vartheta$-functions.
(In the related context of adjoint QCD, this was also noted in Ref.~\cite{Basar:2014jua}.) 
Specifically, the Dedekind $\eta$-function has the product representation 
      $\eta(\tau) = q^{1/24} \prod_{n=1}^{\infty} (1-q^n)$, 
while the generalized $\vartheta$-function
      $\Th{\alpha}{\beta}(\tau) \equiv \sum_{n \in \mathbb{Z}} q^{(n+\alpha)^2/2} e^{2 \pi \ii n \beta}$
has a product representation of the form
\begin{align}
\Th{\alpha}{ \beta}(\tau)
   &=q^{\alpha^2/2}\prod_{n=1}^{\infty}\bigg[(1-q^n) \nonumber\\
   &\times(1+q^{n-\half+\alpha}e^{2 \ii \pi \beta })(1+q^{n-\half-\alpha} e^{-2 \ii \pi \beta }) \bigg]~.~~~~ 
\end{align}

Under the $S:\tau\to -1/\tau$ and $T:\tau\to\tau+1$ generators of the modular group $SL(2,\mathbb{Z})$, 
we find 
$\eta(-1/\tau) =\sqrt{-\ii\tau}\,\eta(\tau)$ and
$\eta(\tau +1) = e^{\ii \pi/12}\, \eta(\tau)$, 
while  
\begin{align}
S: ~~\Th{\alpha}{ \beta}(-1/\tau) 
     &=\sqrt{-\ii \tau}\,e^{-2\pi \ii \alpha \beta}\Th{- \beta}\alpha(\tau)  ~,\nonumber\\
T: ~~\Th{\alpha}{ \beta}(\tau+1)  &=e^{\ii \pi \alpha^2     }\Th\alpha{ \beta+\alpha+1/2}(\tau)~.
\label{eq:orb4}
\end{align}

Given these definitions, the structure of Eq.~\eqref{eq:YM3} allows us to rewrite the 4D partition function
$Z_{\rm YM}$ as a finite product of Dedekind $\eta$-functions and Jacobi $\vartheta$-functions:
\begin{align}
Z_{\rm YM}(\tau)
       % ~&=~  -\sqrt{2}  \,e^{-i\pi b} \, \frac{\eta^5}{\eta(2\tau)} \frac{1}{\Th{1/2}{b+1/2}}  
      %   \nonumber\\  
   ~& =~  \, \, \eta(\tau)^3 \left(\frac{-\sqrt{2} e^{-i\pi b}\eta(\tau)}{\Th{1/2}{b+1/2}(\tau)}\right) \sqrt{\frac{2\,\eta(\tau)}{\vartheta_2(\tau)}}
\label{eq:mod1}
\end{align}
where $b=i \log(z)/2\pi \approx 0.21 i$,
where $\vartheta_{2}(\tau)\equiv \Th{1/2}{0}(\tau)$,
and where the identity 
$2\eta(2\tau)^2 = \eta(\tau) \vartheta_{2}(\tau) $ has been used in passing from Eq.~\eqref{eq:YM3} to Eq.~\eqref{eq:mod1}.  The fact that $b$ is imaginary is the reason the degeneracy factors $d_n$ in Eq.~\eqref{eq:YM1} grow as $d_n \sim e^{Cn}$.
The expression in Eq.~\eqref{eq:mod1} --- and our interpretation of this expression in terms 
of specific 2D CFTs, as discussed below --- are the key results of our paper, with many striking consequences.

%  %%%%%%%%%%%%%%
%  \begin{figure}[ht]
  %  \centering
%  \includegraphics[width=0.47\textwidth]{smallLYM.pdf}
  %  \caption{The small-$|\tau|$ behavior of $Z_{\rm YM}$.}
  %  \label{fig:smallLYM}
%  \end{figure}
%  %%%%%%%%%%%%%%

%%%%%%%%%%%%%%%%%%%%
{\bf Modularity versus dimensionality.} The first interesting implication of Eq.~\eqref{eq:mod1} becomes apparent upon realizing that it is extremely unusual for the partition function of a 4D theory 
to be expressible as a finite product of modular functions, as in Eq.~\eqref{eq:mod1}.  (See, {\it e.g.}\/, 
Ref.~\cite{Cardy:1991kr} for an early discussion along these lines.)    
The large-$|\tau|$  behavior of a modular function 
is tied, through the $S$ modular transformation, to its 
behavior near $|\tau| = 0$.  For example, 
the Dedekind $\eta$-function has the large-$|\tau|$ expansion 
$\eta(\tau) = q^{1/24}(1-q+...)$;  the $S$ transformation then requires 
this function to behave 
at small $|\tau|$ 
as $\eta(\tau) \sim \exp[-i\pi/(12 \tau)]/\sqrt{-i\tau}$. 
Similar statements can be made for the $\vartheta$-functions.  
Thus, if a partition function can be written as a finite product of modular $\eta$-functions and 
$\vartheta$-functions, then it must have the leading behavior
\begin{align}
\lim_{\arg \tau \to \pi/2} \left[\lim_{|\tau| \to 0} \log Z_{\rm modular}(\tau)\right] ~=~ \sigma R/\beta
\label{eq:2DBehavior}
\end{align}
for a constant $\sigma$.  This amounts to the statement that $\log Z_{\rm modular} \sim T$ as $T\to\infty$. 
This is indeed the expected behavior for a 2D QFT.~ 
However, this is certainly {\it not}\/ the expected behavior for a 4D QFT, for which we generically 
expect\footnote{Supersymmetric theories on $S^3 \times S^1$ are an exception: 
     it has been shown that the coefficient of the $T^3$ term in $\log Z_{\rm 4D}$ 
     vanishes at any $N$ if the compactification does not break 
     supersymmetry~\cite{DiPietro:2014bca,*Assel:2015nca}. }
\begin{align}
\log Z_{\rm 4D} \sim T^3 ~~~~ {\rm as}~ T\to\infty~.
\label{eq:4DBehavior}
\end{align}
In this sense, 4D QFTs whose partition functions can be written in terms of modular functions behave as if they were 2D QFTs, since they follow Eq.~\eqref{eq:2DBehavior} rather than Eq.~\eqref{eq:4DBehavior}.   

If we were to reverse the order of limits on the left side of Eq.~\eqref{eq:2DBehavior} and take the $|\tau| \to 0$ limit with $\arg \tau =\pi/2$, pure YM theory would follow the scaling in Eq.~\eqref{eq:4DBehavior}. Such a limit cannot be studied from Eq.~\eqref{eq:YM1} due to the Hagedorn singularities, and the physics is governed by the deconfined phase~\cite{Aharony:2003sx}. For Yang-Mills theory, Eq.~\eqref{eq:2DBehavior} is thus valid only with the order of limits indicated.  We note that in other theories such as adjoint QCD with periodic boundary conditions for fermions, the Hagedorn singularities do not lie along $\arg\tau=\pi/2$~\cite{Basar:2014jua};  the two limits then commute and these theories exhibit 2D behavior in the sense of Eq.~\eqref{eq:2DBehavior} irrespective of the order of limits~\cite{GCDM}.

%For pure Yang-Mills theory, $|\tau| = \frac{\beta}{2\pi R}\sqrt{1+(\mu_{I}R)^2}$ and $\arg \tau = \cot^{-1}(\mu_I R)$, and for Eq.~\eqref{eq:2DBehavior} to be valid it is important that the small $|\tau|$ limit be taken before the $\arg \tau \to \pi/2$ limit, so that we take $L/R \to 0$ before $\mu_I R \to 0$. The limits do not commute because of the Hagedorn singularities found along $\arg \tau = \pi/2$.  If $\arg \tau$ is taken to $\pi/2$ first, then the small $|\tau|$ physics will be governed by the deconfined phase, and cannot be studied from \eqref{eq:YM1}.  In other theories, such as adjoint QCD with periodic boundary conditions for fermions, the two limits commute\cite{Basar:2014jua}.

%%%%%%%%%%%%%%%%%%%%%%%%%%%
{\bf Vacuum energy.}
Another major consequence of Eq.~\eqref{eq:mod1} 
is that the modular properties 
of the $\eta$- and $\vartheta$-functions fix the vacuum energy $E_{\rm YM}$ of our large-$N$ YM theory to be zero.

To see this, we first recall that if we write the $q$-series expansion of a 
%meromorphic 
modular 
function $f(\tau)$ in the form $f = q^{\Delta}\sum_{n = 0}^{\infty} a_n q^n$, 
then $\Delta$ can be thought of as the 2D vacuum energy.  
Its value is fixed by the modular properties of $f$ and tied to the values of $a_n$. 
Were one to abitrarily shift $\Delta \to \Delta + c$,  the modular properties 
of $f(\tau)$ would be ruined because the $S$-transformation would map 
$q^{c} = e^{(2 \pi i \tau) c}$ to $e^{(-2\pi i /\tau)c}$, thereby preventing $q^c f(\tau)$ from transforming 
as a modular function.

Next, we observe that 
the vacuum energy associated to the $\eta$-function is $1/24$, while $\Th{a}{b}$ has vacuum energy $a^2/2$.  Summing the vacuum energies of the individual modular functions in Eq.~\eqref{eq:mod1} we obtain a striking result:
\begin{align} 
E_{\rm YM} ~=~ 0~.
\end{align}
Indeed, this is the only value consistent with the $q$-expansion for $Z_{\rm YM}$ given in Eq.~\eqref{eq:YM1}, provided that $E_{\rm YM}$ is calculated in a renormalization scheme which is consistent with the modular 
properties of $Z_{\rm YM}$ made evident in Eq.~\eqref{eq:mod1}. 
This value, $E_{\rm YM} = 0$, coincides with the result implied by T-reflection symmetry~\cite{Basar:2014mha}, and also agrees with a direct evaluation of the sum over the confined-phase spectrum of finite-temperature large-$N$ YM theory compactified on $S^3$, as performed in Ref.~\cite{Basar:2014hda}.

%%%%%%%%%%%%%%%%%%%%%%%%%%%
{\bf CFT interpretation.}  
The striking modular structure of Eq.~\eqref{eq:mod1} suggests that the spectrum of our 4D YM theory 
coincides with that of a chiral ({\it e.g.}\/, left- or right-moving) 2D CFT.~
%Indeed, 
%if we rewrite $\eta(\tau)^3 = \eta(\tau)^2 \eta(\tau)^2 \eta(\tau)^{-1}$, 
%each of the multiplicative factors within the final expression in
%Eq.~\eqref{eq:mod1} can be interpreted as the trace over the Fock space of left-moving states of a 2D CFT,
%so that the entire expression can be viewed as the trace over Fock space of chiral states of a single direct-product CFT.~    
This motivates the central question we shall now explore for the rest of this paper: what is the 2D CFT which gives rise to Eq.~\eqref{eq:mod1}, and thus gives a 2D description of 4D YM theory in the large-$N$ limit?

Unfortunately, we will not be able to give a
complete answer to this question.
The reason ultimately has to do with the fact that
many distinct CFTs can have coincident spectra without being equivalent.
They may differ, for example, in their correlation  functions.  
In general, the most important aspects of a given 2D CFT are governed by 
its central charge (conformal anomaly) $c$ and its spectrum of operator conformal dimensions $h_i,\, i=1,...,n$, where
$n$ is the number of so-called ``primary'' fields in the CFT.~
Along with the explicit traces over states, 
knowledge of $c$ and the $h_i$'s goes a long way in nailing down relevant aspects of the CFT such
as its selection rules and correlation functions.
But partition functions are only sensitive to the combinations $h_i^{\rm (eff)}\equiv h_i-c/24$, rather than the 
values of $c$ and $h_i$ individually.
Consequently, without additional assumptions about the CFT in question (such as the assumption of unitarity, which  
would additionally tell us that $\mathrm{min}\, \{h_i\} = 0$), this represents a fundamental limitation on our ability 
to specify a unique CFT.   

We will therefore answer a different but related question: 
do there exist any 2D CFTs to which our large-$N$ YM theory is {\it isospectral}\/?  
Remarkably, we shall show that at least one such 2D CFT indeed exists.
To see this, we first
recall that a free $c=1$ scalar CFT has a chiral spectrum whose trace is given by $1/\eta(\tau)$, 
while the $\mathbb{Z}_2$ orbifold of this CFT has a chiral sector whose trace is
$(2\eta(\tau)/\vartheta_2(\tau))^{1/2}$.  
Furthermore, the direct product of two copies of the $c = -26$ 
$bc$ ghost CFT has a chiral spectrum whose trace is given by $\eta(\tau)^4$.  
Perhaps the most challenging to interpret is the remaining
factor in  Eq.~\eqref{eq:mod1}, specifically
\begin{align}
\frac{-\sqrt{2} e^{-i\pi b}\eta(\tau)}{\Th{1/2}{b+1/2}(\tau)}~.
\end{align}
However, this can be identified as the trace of the chiral ({\it e.g.}\/, left-moving) 
states in the vacuum sector 
of the $c=2$  bosonic $\beta \gamma$ ghost CFT recently explored in Ref.~\cite{Ridout:2014oca}.  
This is a logarithmic CFT~\cite{Gurarie:1993xq}, and it has a $U(1)$ conserved charge.  
Thus the vacuum-sector chiral partition function of the $c=2$ $\beta \gamma$ CFT depends on the choice of a complex fugacity $z = e^{+\mu\beta}$.  
To match with our expressions for YM theory, we set  $\mu\beta = 2\pi \ii \, b =- \log(2+\sqrt{3})$.

Putting this together, we therefore conclude that the expression in Eq.~\eqref{eq:mod1} can 
be viewed as the trace over the chiral spectrum of a theory which is the direct product of five known CFTs, 
one of which is irrational.  This then justifies 
the central claim of this paper in 
Eq.~\eqref{eq:TheClaimYM}: 
there is indeed an irrational 2D CFT which is isospectral 
to the finite-temperature large-$N$ 4D YM compactified on $S^3$ in the $\Lambda R \to 0$ limit.   Aside from explaining
our observations concerning $E_{\rm YM}$ and the small-$|\tau|$ behavior of $Z_{\rm YM}$, 
this fact has an intriguing further implication.  Two-dimensional CFTs have infinite-dimensional symmetries, which always include the Virasoro symmetry.  Eq.~\eqref{eq:TheClaimYM} then suggests that large-$N$ YM theory has a hidden Virasoro symmetry.  It would be very interesting to demonstrate this explicitly within YM theory.

%%%%%%%%%%%%%%%%%%%%%%%%%%%
{\bf Primary operator spectrum.}
We now collect information concerning the spectrum of conformal dimensions $h_i^{\rm (eff)}$ corresponding to the
primary fields of this tensor-product CFT.~
Our approach proceeds by 
determining the diagonal modular-invariant associated with the expression in Eq.~\eqref{eq:mod1}, and then computing the eigenvalues of the modular $T$ operator to extract $h_i^{\rm (eff)}$.  

We begin by defining the quantities
\begin{align}
    T_{m,n} ~\equiv~ \frac{-\sqrt{2} \, e^{-i\pi b n}\, \eta(\tau)^4}{\Th{mb+1/2}{nb + 1/2}(\tau)}  \left(\frac{2\eta(\tau)}{\Th{P(m)/2}{P(n)/2}(\tau)}\right)^{1/2} ~, 
\label{eq:orb5}
\end{align}
where $\{m,n\}$ are relatively prime integers (a relationship which we shall henceforth denote $m\perp n$), 
and  $P(k) \equiv  \frac{1}{2}(1+(-1)^k), k \in \mathbb{Z}$. 
Thus $P(k)=0,1$ for odd or even $k$, respectively.
 The set $\{T_{m,n}\}$ is a basis for a vector space over the field $\mathbb{C}$ with two key properties:  it contains the ``seed term'' in Eq.~\eqref{eq:mod1}, and it is the minimal set which is closed under the action of the $SL(2, \mathbb{Z})$ modular group.

The first property follows by noting that $T_{0,1}(\tau)$ coincides with Eq.~\eqref{eq:mod1}.
The verification of the second property proceeds in two steps. 
First, it can be shown that, 
up to overall phases and extraneous factors of $\sqrt{-i\tau}$, 
the $S$ and $T$ modular transformations map $T_{m,n}$ to $T_{-n,m}$ and $T_{m,n+m}$, respectively.  
Second, we observe that if $\{m,n\}$ are relatively prime, then $\{-n,m\}$ and $\{m,n+m\}$ are also relatively prime.
Since all modular transformations can be generated by sequences of $S$ and $T$,
it then follows that the full modular ``orbit'' 
of our seed term $T_{0,1}$ is contained within the set of coprime integers $\{m,n\}$.
Indeed, it is also possible to demonstrate~\cite{GCDM} that the modular 
orbit actually covers {\it all}\/ coprimes.

%%%%%%%%%%%%%%
\begin{figure}[ht]
  \centering
\includegraphics[width=0.47\textwidth]{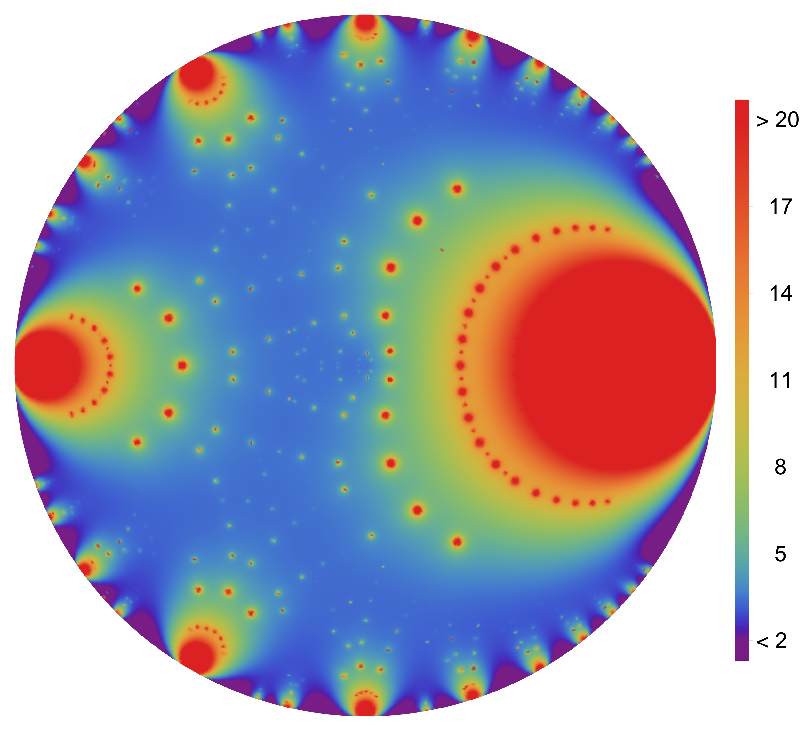}
  \caption{The numerical values of Eq.~\eqref{eq:orb6} with $|m|,|n|\leq 10$, plotted within the unit-$q$ disk.}
\label{fig:YMmodularity}
\end{figure}
%%%%%%%%%%%%%%

As a result, the minimal ``diagonal'' modular-invariant generated from Eq.~\eqref{eq:mod1} is given by 
\begin{align}
   Z_{\rm diagonal} ~=~ ({\rm Im}\, \tau)^{3/2} 
\sum_{m \perp n}  \big| T_{m,n} \big|^2  ~ . 
\label{eq:orb6}
\end{align}
The appearance of the factor of $({\rm Im}\,\tau)^{3/2}$ is standard when combining holomorphic and anti-holomorphic components,
such as our $T_{m,n}$ factors,
each of which has modular weight $k=3/2$.  It also ensures
that $Z_{\rm diagonal}$ is fully modular-invariant.
Moreover, it can be verified numerically that the infinite sum in Eq.~\eqref{eq:orb6} converges except for an isolated set of points corresponding to the Hagedorn singularities.
%, as required
%Fig.~\ref{fig:YMmodularity} contains a plot of $Z_{\rm diagonal}$ on the interior of the unit-$q$ disk.
The numerical values of $Z_{\rm diagonal}$
on the interior of the unit-$q$ disk are shown in 
Fig.~\ref{fig:YMmodularity}.

In order to extract the spectrum of effective conformal dimensions $h^{\rm (eff)}_i$,
we now rewrite $Z_{\rm diagonal}$ in a basis of eigenfunctions of the modular $T:\,\tau\to\tau+1$ operator. 
%  (Similar calculations are performed, {\it e.g.}\/, in Sect.~4 of Ref.~\cite{Dienes:1994np}.)
We do this because such eigenfunctions $\chi(\tau)$ will have eigenvalues $\exp[2\pi i h_i^{\rm (eff)}]$ under $T$,
allowing us to read off the values of $h_i^{\rm (eff)}~({\rm mod}~1)$.
%  (Determining the actual values of $h_i^{\rm (eff)}$, by contrast, would require an explicit 
%  $q$-expansion of $\chi(\tau)$, which in our case will be more difficult to obtain.) 
Fortunately, constructing eigenfunctions of the $T$-operator from linear combinations of the $T_{m,n}$'s 
in Eq.~\eqref{eq:orb5} is relatively straightforward.
Since 
\begin{align}
   T_{m,n}(\tau+1) = 
   e^{\pi i \left\lbrace  [1-P(m)]/8 +m^2 |b|^2  \right\rbrace} \,T_{m,n+m}(\tau) ~ , 
\label{eq:char1}
\end{align}
we see that any linear combination which includes $T_{m,n}$ must also include $T_{m,n+m}$,
$T_{m,n+2m}$, and indeed all $T_{m,n+km}$ where $k\in\mathbb{Z}$.
Our $T$-invariant linear combinations can therefore be indexed by an arbitrary integer $m$
and a second integer $\ell \perp m$ obeying $0 \leq \ell < |m|$.
%In general, for any $m$, there will be $\phi(m)$ different choices of $\ell$, where $\phi(m)$ is the Euler totient
%function.
Hence $T$-eigenfunctions can be constructed analogously to Bloch eigenfunctions,  by summing over all
components $T_{m,\ell+k m}$ with $k\in \mathbb{Z}$ with a Bloch phase $\alpha \in [0,1) \subset \mathbb{R}$:
\begin{align}
\chi_{m,\ell, \alpha} ~=~\sum_{k\in \mathbb{Z}} e^{2 \pi \ii \alpha k} \, T_{m,\ell+ m k}~.
\label{eq:char2}
\end{align}
It then follows that
\begin{align}
\chi_{m,\ell, \alpha}(\tau+1) ~=~ e^{2\pi \ii h^{\rm (eff)}_{m,\ell,\alpha}} \, \chi_{m,\ell, \alpha}(\tau)~,
\end{align}
where 
\begin{align}
     h^{\rm (eff)}_{m,\ell,\alpha} ~=~ \frac{1}{2}\left[ \frac{1- P(m)}{8} + m^2|b|^2\right] - \alpha~.
\label{eq:hvalues}
\end{align}

One might wonder whether  $\{\chi_{m,\ell,\alpha}\}$ is the complete set of $T$-eigenfunctions. However, we 
have verified this by checking that summing over $\chi_{m,\ell,\alpha}$
reproduces Eq.~\eqref{eq:orb6}:
\begin{align}
 Z_{\rm diagonal} = ({\rm Im}\, \tau)^{3/2}\, 
\sum_{m \in \mathbb{Z}}\, \sum_{\substack{
          0\leq \ell < |m|\\
            \ell\perp m}} \, \int_{0}^{1} d\alpha \, |\chi_{m,\ell, \alpha}|^2 .
\end{align}

This confirms that Eq.~\eqref{eq:hvalues} is the desired set of effective conformal dimensions (mod $1$) of the primary operators in our CFT.~
The fact that these dimensions depend on $\alpha$ --- a continuous real variable --- confirms that
we are dealing with an {\it irrational}\/ CFT~\cite{Anderson:1987ge,*Vafa:1988ag}.  
Our observations are consistent with the  2D logarithmic CFT interpretation discussed above, 
since it is known that logarithmic CFTs typically have a continuously infinite 
number of primary operators~\cite{Creutzig:2013hma}.

{\bf Outlook.} We have presented evidence 
that the confined phase of finite-temperature 4D non-supersymmetric 
large-$N$ pure Yang-Mills theory compactified on a three-sphere of radius $R$ 
has a remarkable modular structure, as exposed by Eq.~\eqref{eq:mod1}.  This has many interesting consequences, such as the fact that this 4D gauge theory is isospectral to an irrational 2D CFT in the $\Lambda R\to 0$ limit, as summarized in Eq.~\eqref{eq:TheClaimYM}.   
%This gives credence to the hope and expectation that non-supersymmetric large-$N$ confining gauge theories are dual to 2D CFTs. 
Moreover, as we shall demonstrate in a separate paper~\cite{GCDM}, 
modularity in the sense of Eq.~\eqref{eq:mod1} 
and isospectrality to 2D irrational CFTs as in Eq.~\eqref{eq:TheClaimYM}
turn out to be  generic properties of large-$N$ confined-phase gauge theories with adjoint massless matter 
in the $\lambda\to 0$ limit.  In Ref.~\cite{GCDM} we shall also show that this structure is present in the large-$N$ limit of the $\mathcal{N}=4$ superconformal index. 

It is not clear how easily the 4D-2D relation that we found, as summarized through Eq.~\eqref{eq:TheClaimYM}, 
fits with standard string-theoretic expectations.  From a string-theoretic perspective one might have 
expected that it would be the {\it single-trace}\/ partition function --- which can be thought of 
as representing the fluctuations of a single string --- that would have a simple 2D CFT description, assuming one is possible.  It is less clear why the grand-canonical partition function $Z_{\rm YM}$, which takes into account all multi-trace states and hence represents the fluctuations of an ensemble of many strings, should have a 2D CFT description.  From this perspective, our result in Eq.~\eqref{eq:TheClaimYM} --- and the analogous relations that we shall find in Ref.~\cite{GCDM} for other, adjoint-matter gauge theories --- are even more remarkable. 

Our results suggest a large number of interesting topics for future research.   Obviously, it would be very interesting to understand whether Eq.~\eqref{eq:TheClaimYM} has an explanation within string theory, perhaps by making contact with the ideas in, {\it e.g.}\/, Refs.~\cite{Gopakumar:2003ns,*Gopakumar:2004qb,*Itzhaki:2004te,*Razamat:2012uv,*Cordova:2015nma}.  It is also important to understand whether our large-$N$ 4D-2D spectral equivalence extends to correlation functions, and to explore how it is related to other known 
4D-2D relations, such as those discussed in Refs.~\cite{Alday:2009aq,*Beem:2013sza}.  Note that unlike the 4D-2D relations discussed in the context of supersymmetric indices (which by construction focus on a subset of states of the 4D theory), our 4D-2D relation in Eq.~\eqref{eq:TheClaimYM} involves the full thermal partition function and hence concerns the entire spectrum of the 4D theory.  

Another interesting direction would be to develop an understanding of the modular structure of expressions like Eq.~\eqref{eq:mod1} directly from a 4D point of view, perhaps by making use of ideas from, {\it e.g.}\/, 
Refs.~\cite{Closset:2013vra,*Nieri:2015yia}.
Given recent progress in the understanding of the bulk duals of 2D CFTs~(see, {\it e.g.}\/, Ref.~\cite{Gaberdiel:2012uj}), it is tempting to wonder whether our results may help to uncover the bulk dual of YM theory and of other non-supersymmetric 4D adjoint-matter theories. It would also be interesting to understand the extent to which the continuous spectrum of primary operators in the 2D theory suggested by our analysis has an interpretation in 4D YM theory.  
Finally, there remains the very important question of determining how the modularity of $Z_{\rm YM}$ and our ensuing 4D-2D relation might evolve for $\lambda >0$.  

%%%%%%%%%%%%%%%%

{\bf Acknowledgements.}  We are grateful to O.~Aharony, D.~Berenstein, S. Caron-Huot, S.~Cremonesi, N.~Dorey, G.~Dunne, D.~Hofman, J.~Kaplan, Z.~Komargodski, P.~Koroteev, J.~McGreevy, D.~O'Connor, E.~Poppitz, S.~J.~Rey, D.~Ridout, A.~Shapere, D.~Tong, M.~Unsal, S.~Wood, and M.~Yamazaki for helpful discussions at various stages during the gestation of this work.  We are especially grateful to Z.~Komargodski for helpful comments on an earlier version of this paper.  
AC thanks the Instituto de Fisica Teoretica of UAM and DAMTP of Cambridge University for hospitality, and DAM thanks the Institute for Advanced Study and the Niels Bohr International Academy for hospitality and support during the completion of this work.  GB and AC thank the Department of Energy for support under Grants DE-FG02-93ER-40762 and DE-FG02-94ER40823 respectively.
The research of KRD was supported in part by the Department of Energy
under Grant DE-FG02-13ER-41976 and by the National Science Foundation through its employee IR/D program.
The opinions and conclusions expressed herein are those of the authors,
and do not represent any funding agencies.

\bibliographystyle{apsrev4-1}
\bibliography{super_susy} 
%%%%%%%%%%%%%%%%%%%%%%%%%%%%%%%%%%%%%%%%%%%%%%%%%%%%%%%%%%%%%%%%%%%%%%%%%%%%%%%%%%%%%%%%%%%%%%%%%%%%%%%%%%%%%%%%%%%%%%%%%%%%%%%%%%%%%%%%%%%%%%%%%%%%%%%%%%%%%%%%%%%%%%%%%%%%%%%%%%%%%%%%%%%%%%%%%%%%%%%%%%%%%%%%%%%%%%%%%%%%%%%%%%%%%%%%
\end{document}